\documentclass[pre,aps]{revtex4}
\usepackage{amssymb}
\usepackage{graphicx}
\usepackage{amsmath}
\usepackage{times}
\newcommand{\be}{\begin{equation}}
\newcommand{\ee}{\end{equation}}
\newcommand{\ab}{{\mathcal A}}  

\begin{document}
\title{Equilibrium state of molecular breeding}
\author{Elisheva Cohen}
\affiliation{Dept. of Physics, Bar-Ilan University, Ramat-Gan IL52900 ISRAEL}
\author{David A. Kessler}
\email{kessler@dave.ph.biu.ac.il}
\affiliation{Dept. of Physics, Bar-Ilan University, Ramat-Gan IL52900 ISRAEL}

\begin{abstract}
We investigate the equilibrium state of the model of Peng, \textit{et al.}
for molecular breeding.  In the model, a population of DNA sequences
is successively culled by removing the sequences with the lowest
binding affinity to a particular target sequence.  The remaining
sequences are then amplified to restore the original population size,
undergoing some degree of point-substitution of nucleotides in the process.
Working in the infinite population size limit, we derive an equation for 
the equilibrium distribution of binding affinity, here modeled by the number
of matches to the target sequence. The equation is then solved
approximately in the limit of large sequence length, in the three
regimes of strong, intermediate and weak selection.  The approximate
solutions are verified via comparison to exact numerical results.
\end{abstract}
\maketitle

\section{Introduction}
Recent advances in molecular biology have enabled breeding 
methodologies to be applied at the molecular level to develop novel
proteins and DNA sequences with particular desired characteristics\cite{Farinas}.
The basic idea is simple: in each generation, a number of 
molecules are selected from the population according to some criteria
of desirability; they are then diversified (via point mutation and/or
recombination) and then amplified back to the original population size.
This methodology is worthy of study not only for its immediate practical
importance but also as a forum for exploring general issues in evolution 
and population dynamics in a uniquely controllable setting.

Recently, motivated in particular by the experiment of Dubertret, {\em et al.}
\cite{dubertret} on the {\em in vitro} evolution of DNA sequences,
Peng, {\em et al.} (PGHL) \cite{PGHL} introduced a simple model of molecular evolution.
In the experiment, at each round a specified percentage of the population
was selected to survive based on its binding affinity to a particular
target, the {\em lac}-repressor protein.  This was accomplished by coating
a beaker with the {\em lac}-repressor protein into which the DNA was 
introduced. The beaker was washed out, so that only the most strongly bound
sequences remained.  The degree of selection could be directly controlled
by the strength of the washing procedure.  Amplification, accompanied
by mutation, was subsequently performed by multiple stages of PCR.
The model of PGHL incorporates these basic processes of selection,
mutation and reproduction.  Each sequence is modeled as a string of
$L$ nucleotides, each taken from the set of $\ab=4$ ``letters'', \{A,G,C,T\}.  
The binding affinity of the sequence is taken to be
proportional to the number of nucleotide matches between it and the
target sequence. The fraction $(1-\phi)$ with the lowest affinities
are dropped from the population.  The population is then amplified back
to its original level, where each daughter sequence in the next round is 
chosen to be descended from one of the surviving sequences, modified by
point mutations at a rate of $\mu_0$ per nucleotide.

PGHL performed simulations on their model and found that, starting from
an initial population with low affinity, the average affinity of the
population rose, eventually reaching equilibrium.  They then constructed
a mean-field (i.e. infinite population) treatment of the problem.  They
showed that the average affinity relaxes to its equilibrium value
exponentially in time, with a time constant of $\ab\mu_0/(\ab-1)$,
and verified this via simulation.  However, the calculation of
the equilibrium value itself was less successful.  In their treatment,
this value was found to depend on a parameter which was introduced
in an ad-hoc manner into their theory, and which needed to be fit
from simulation.  In this paper, we reexamine the equilibrium problem,
locating the source of the difficulty of the PGHL treatment in an 
inappropriate continuum approximation.  In turns out that the
problem is fairly subtle, and in fact requires different treatment in
the case of very weak, very strong and intermediate selection.  The
range of the validity of the intermediate selection increases with
$L$, so that for sufficiently large $L$, the average affinity is
given for any $\phi$ by the intermediate result.

The plan of the paper is as follows.  In Section \ref{basics}, we lay
out the general formalism and basic equations that we need to solve.
In Sec. \ref{strong}, we solve the strong selection problem,
working in the small $\mu_0$ limit.  
In Sec. \ref{inter}, we treat the case of 
intermediate strength selection, also for small $\mu_0$.
In Sec. \ref{inter1}, we generalize the intermediate strength solution
to the case of arbitrary $\mu_0$, obtaining the central result of the
paper, the average density of mismatches for any $\phi$ and $\mu_0$,
given that $L$ is sufficiently large.  In Section \ref{weak}, we tackle the
weak selection limit, reverting back to small $\mu$.
In Sec. \ref{inter2}, we return again to the case of intermediate
selection, calculating the next order corrections.  
We discuss the connection to the PGHL treatment in Sec. \ref{compare}.
We conclude in Sec. \ref{end} with some observations.

\section{Basic Formalism}
\label{basics}

We are interested in this paper in calculating the equilibrium distribution
of binding affinity in the population of sequences.  Since the
dynamics only depends on the binding affinity, determined by the
number of matches between the sequence and the target, we can 
characterize each sequence by this quantity, or equivalently, the
number of mismatches, $0 \le k \le L$, in the length $L$ sequence.  We
will work in the infinite population limit, so that fluctuations are
irrelevant, and focus on the master equation for the normalized mismatch
distribution, $P_k$.  We need to track the changes induced in this distribution
by each of the two processes, selection and mutation; amplification in and of
itself having no effect on the distribution.

Selection of the fraction $\phi$ of the highest affinity sequences
throws out those states with the highest $k$'s, keeping the lower $k$
states.  Thus, after selection, some set of states $0 \le k \le n$
are retained in their entirety, along with some fraction $\alpha$ of the
$k=n+1$ state.  The renormalized distribution after selection, $P^s$, is 
given by
\begin{equation}
P_k^s = \left\{ \begin{array}{l c c}
\frac{1}{\phi} P_k & \quad & 0 \le k  \le n \\
\frac{\alpha}{\phi} P_{n+1} & \quad & k = n+1 \\
0 & \quad &  k > n+1 
\end{array} \right.
\ee
where the equation
\be
\phi = \alpha P_{n+1} + \sum_{k=0}^n P_k 
\ee
implicitly determines $n$ and $\alpha$.  

If the mutation rate is sufficiently small,
we need only consider single nucleotide mutation events, so that only 
transitions $k \to k\pm 1$ are allowed.  We will take up the case of larger
mutation rates in Sec. \ref{inter2}. The distribution after mutation, $P^\mu$,
is given by
\be
P^\mu_k = \left(1-\mu+\mu(\bar{\nu})\frac{k}{L}\right)P^s_k + \mu \left(\frac{L-k+1}{L}\right)P^s_{k-1}
+ \mu\nu \left(\frac{k+1}{L}\right) P^s_{k+1} ,
\ee
where $\mu\equiv \mu_0 L$ is the genomic mutation rate, 
$\nu\equiv 1/(\ab-1)$, $\bar{nu}\equiv 1-\nu$, 
and we have set $P^s_{-1}\equiv 0$. 
Mutation repopulates the $k=n+2$ state,
which was emptied by selection.  At equilibrium, the post-mutation distribution
$P^\mu_k$ must reproduce the initial distribution $P_k$, giving us a closed
set of equations for $P_k$. These equations involve only the finite set of
states $0 \le k \le n+2$.  In fact, $P_{n+2}$ is not involved in any
of the equations, since its initial value is zeroed out by selection.  Thus,
the equilibrium equations close on the $n+2$ quantities $P_k$, $0 \le k \le n+1$, with $P_{n+2}$ being slaved to the solution for $P_{n+1}$.  This
system is ($0 \le k \le n+1$)
\begin{eqnarray}
P^\mu_k &=& \frac{1}{\phi}\left[ \left(1-\mu+\mu\bar{\nu}\frac{k}{L}\right)f_k P_k + \mu \left(\frac{L-k+1}{L}\right) f_{k-1}P_{k-1} \right . \nonumber \\
&\ & \quad \left. {}+ \mu\nu \left(\frac{k+1}{L}\right)f_{k+1}P_{k+1}\right] 
\label{exacteq}
\end{eqnarray}
with the filling factor $f_k=1$ for $0\le k \le n$, $f_{n+1}=\alpha$
and $f_{-1}=f_{n+2}=0$.
These equations
for the $P$'s, with a given $n$ and $\alpha$ are linear and homogeneous, with 
$\phi$ playing the role of an eigenvalue condition for the existence of a
nontrivial solution. As $\alpha$ increases for a given $n$, $\phi$ increases,
implying weaker selection, till $\alpha$ reaches its limiting value of $1$.
At this point, we increase $n$ and start $\alpha$ again at $0$.  Thus the
solution is piecewise analytic in $\phi$. This piecewise behavior is seen
clearly in Fig. (\ref{reg1-50}), where the numerical solution for $L=50$,
$\mu=0.1$, $\ab=2$ is presented.

For simplicity here, we will only treat the case of $\alpha=1^-$, where
the last selected level is in fact completely selected.  Writing $N=n+2$
and setting $P_{N}=0$ for convenience (its true value being determined
afterwards), we can write the equations in their final form
\be
\chi P^\mu_k = \bar{\nu}\frac{k}{L}P_k +  \left(\frac{L-k+1}{L}\right)P_{k-1}
+ \nu \left(\frac{k+1}{L}\right)P_{k+1} \quad \quad
(0 \le k \le N-1 ) .
\label{basiceq}
\ee
with the boundary conditions $P_{-1}=P_N=0$, 
where we have introduced the scaled eigenvalue
\be
\chi\equiv \frac{\phi - 1 + \mu}{\mu}  .
\ee
After the solution of this system, the $P$'s can be normalized via
$\phi=\sum_{k=0}^{N-1}P_k$, and then $P_N$ can be constructed from $P_{N-1}$ via
\be
P_N = \frac{\mu P_{N-1}}{\phi}\left(\frac{L-N+1}{L}\right) .
\ee

We see that the system equilibrium is controlled by the single combined 
parameter $\chi$,
rather than the two parameters $\phi$ and $\mu$ separately.  Increasing
$\phi$ or $\mu$ both act to increase $\chi$, and as we shall see, increase
the average mismatch.  Since the rate of approach to equilibrium is governed
solely by $\mu$, one can achieve the same equilibrium in less time by
increasing $\mu$ and the selection strength (thereby decreasing $\phi$)
simultaneously.

We see that there is a special role to the value $\phi=1-\mu$, where
$\chi=0$.  This
is the limit of maximal selection, since at this point only the $k=0$
is selected.  Lower values of $\phi$ just result in a partial selection
of the $k=0$ state, which changes nothing since, after amplification, 
the result is the same. Thus, for $\phi \le 1-\mu$ the solution is
$P_0=1-\mu$, $P_1=\mu$, and the mean mismatch $\bar{k}=\mu$.  The other
limit of $\chi$ is given by the weak selection limit $\phi \to 1^-$, where
$\chi$ also approaches $1$.

The system of equations Eq. (\ref{basiceq}) is not analytically solvable
in general, but simplifies in the limit of large $L$, which we now
proceed to study.

\section{Strong Selection, $N\ll L$}
\label{strong}

We first consider the case of strong selection, where the number of
selected states, $N$, is much less than the sequence length $L$.
The system Eq. (\ref{basiceq}) can then be written as

{\small
\begin{equation}
\left( \begin{array}{c c c c c c c}
\chi  &  -\frac{1}{L}           &                 &             & & & \\
-1    &\,\chi - \frac{\bar{\nu}}{L}\, & -\frac{2\nu}{L} &             & & & \\
      &\frac{1-L}{L} &\,\chi - \frac{2\bar{\nu}}{L}\,  & -\frac{3\nu}{L}    & & & \\ 
             &              &                          & \ddots      & & & \\
& & & & \frac{N-3-L}{L}  &\chi - \frac{(N-2)\bar{\nu}}{L} & -\frac{(N-1)\nu}{L} \\
& & & &                  &\frac{N-2-L}{L}       &\chi - \frac{(N-1)\bar{\nu}}{L} 
\end{array} \right)
\left( \begin{array}{crcr}P_0\\ P_1\\ P_2\\\vdots\\ P_{N-2}\\P_{N-1}\end{array} \right)
=\left ( \begin{array}{crcr}0\\ 0\\ 0 \\
\vdots \\0 \\0 \end{array} \right)
\label{matrix}
\end{equation}}

This homogeneous set of equations has a solution when the determinant,
${\cal D_N}$ vanishes,
providing an eigenvalue condition on $\chi$.  Expanding in 
minors about the last column gives a recursion relation 
\begin{equation} 
{\cal D}_N=\left(\chi- \frac{\bar{\nu}(N-1)}{L}\right) {\cal D}_{N-1} 
+ \left(\frac{(N-1)\nu}{L}\right)\cdot
\left(\frac{N-2}{L}-1\right){\cal D}_{N-2} \ .
\label{det}
\end{equation}
Dropping the $\frac{\bar{\nu}(N-1)}{L}$ in the first term and the
$\frac{N-2}{L}$ 
in the second term of the right side of Eq.
(\ref{det}) (because $N\ll L$) gives the equation
\begin{equation}
\tilde{\cal D}_{N+1}=y \tilde{\cal D}_N -N\tilde{\cal D}_{N-1}
\label{y}
\end{equation}
where $\tilde{\cal D}_N=(L/\nu)^{\frac{N}{2}}{\cal D}_N$ and 
$y=\chi\sqrt{L/\nu}$. The solution of Eq. (\ref{y}) is the $N^{\textit{th}}$
Hermite polynomial
$\textrm{He}_N(y)=2^{-\frac{N}{2}}H_N(\frac{y}{\sqrt{2}})$. 
We are thus interested in the zeroes of $\textrm{He}_N(y)$.
In fact, we have to take the maximum root of $\textrm{He}_N(y)$,  which we call
$y^*_N$, because any 
other solution give rise to negative $P_n$'s. Using Maple, for example, 
to compute the $y^*_N$'s, we obtain our approximate solution for
$N \ll L$:
\begin{equation}
\phi_N=1-\mu + \frac{\mu\sqrt{\nu}}{\sqrt{L}}y^*_N
\end{equation}

We now can compute $\bar{k}$, the average
number of mismatches.  Going back to Eq. (\ref{matrix}),
 one can see that $P_1\sim\sqrt{L}P_0$, $P_2\sim\sqrt{L}P_1$ etc., so that 
$P_{N-1}$ is the single dominant term and is thus 1 to leading order ($P_N$ is
of order $\mu$, and so is also irrelevant in our small $\mu$ limit). 
This gives us $\bar{k}\approx N-1$, so that
\begin{equation}
\phi=1-\mu + \frac{\mu\sqrt{\nu}}{\sqrt{L}}y^*_{\bar{k} + 1}
\label{1st_ans}
\end{equation}
This is presented in Fig. 1, along with the exact numerical results
for the case $L=50$, for the case $\ab=2$ ($\nu=1$).  
As we noted earlier, the numerical results are piecewise continuous, due
to the partially filled levels, with the transitions
between levels apparent as cusps in the curve.
We see that the agreement is quite good for the
smaller $N$, and gets worse as $N$ increases, as expected.
\begin{figure}
\centerline{
\includegraphics[width=4.0in]{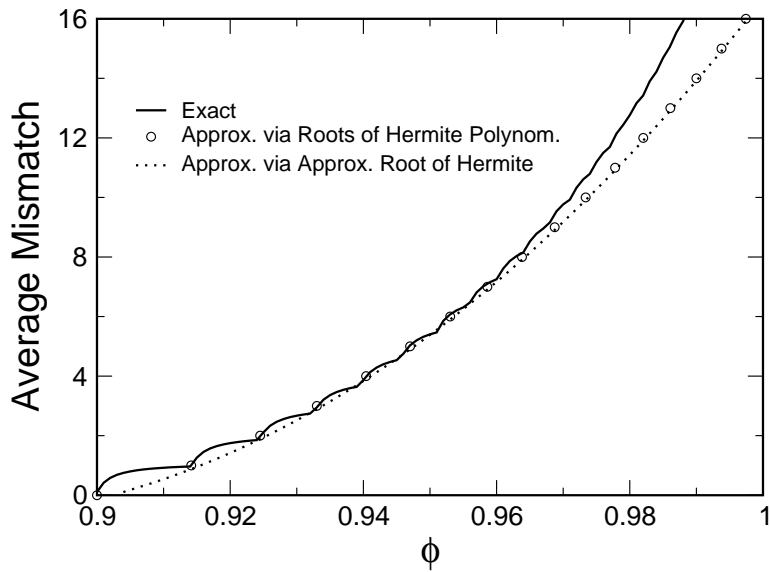}}
\caption{Comparison between numerical results for $L=50$, $\mu=0.1$, $\ab=2$
for the average mismatch, $\bar{k}$, vs. the selection parameter $\phi$ 
(solid line) and our approximation 
$\phi=1-\mu+y_{\bar{k}+1}^*\mu L^{-1/2}$ (circles), 
together with the result using
the asymptotic approximation for $y_N^*$, Eq. (\protect{\ref{herm-approx}})}
\label{reg1-50}
\end{figure}

It is interesting to investigate the behavior of $y^*_N$ for large $N$.
We remember that the $N$'th Hermite polynomial, times a Gaussian, is the 
solution of 
the Schroedinger equation for 
the harmonic oscillator, with $\omega=1/2$ and energy level 
$E_N=(N+\frac{1}{2})\omega$, 
($\hbar=m=1$):
\begin{equation}
-\frac{1}{2}\psi'' + \frac{1}{8}x^2\psi = \frac{1}{2}(N + \frac{1}{2})\psi
\end{equation}
The zeroes of $\textrm{He}_N$ are the zeros of the wave function $\psi$.
For large $N$, the maximal zero is near the classical turning point, where 
the energy equals the potential, so that $x^2/8\approx N/2$, or $x=2\sqrt{N}$.
To achieve a more accurate approximation, we expand the potential around
the turning point, writing
$x=2\sqrt{N}+a(z-z_0)$, and dropping the quadratic term gives an Airy equation
\begin{equation}
\frac{d^2\psi}{dz^2} = z\psi
\end{equation}
if we choose $a=N^{-1/6}$, $z_0=-N^{-1/3}/2$.  The first zero of the
Airy function is at $z=-2.338$, so that
\begin{equation}
\label{herm-approx}
y^*_N=2\sqrt{N} - 2.338N^{-1/6} +\frac{1}{2}N^{-1/2}
\end{equation}. 
The next order correction due
to the dropped quadratic piece can be seen to be of order $N^{-2/3}$.
This is an excellent approximation to $y^*_N$ even for $N=2$, as can be seen
in Fig. \ref{reg1-50}, where this approximation is used
to achieve a completely analytic expression for $\phi(\bar{k})$.

The upshot of all this is that in the first region the scaling variables
are $y=\sqrt{L/\nu}(\phi-(1-\mu))/\mu$ and $\bar{k}$.  Plotting the results
for different $L$ in these variables demonstrates the collapse
beautifully, see Fig. \ref{strongcollapse}.
\begin{figure}
\centerline{
\includegraphics[width=4.0in]{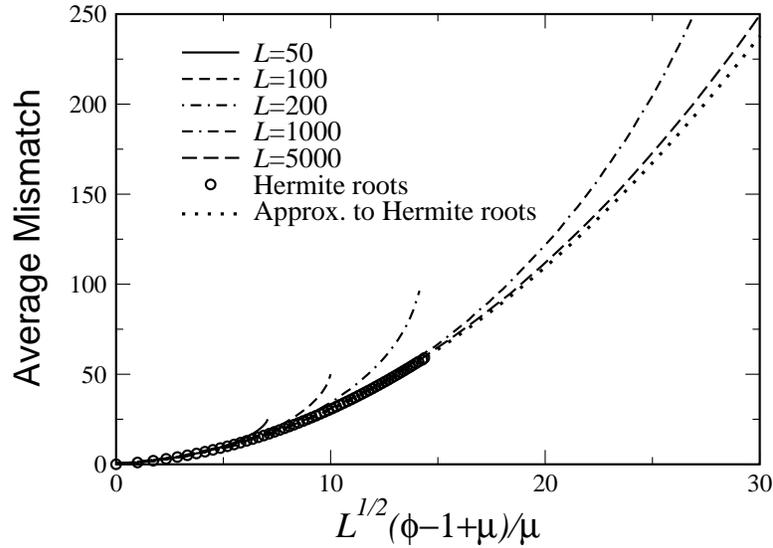}}
\caption{Results for $\mu=0.1$, $\ab=2$, and 
$L=50$, $100$, $200$, $1000$, and $5000$, 
plotted
using the scaling variables $y$ and $\bar{q}$, together with the
first-region approximation, Eq. \protect{\ref{1st_ans}} and
the further approximation to this using the Hermite asymptotics,
Eq. \protect{\ref{herm-approx}}} 
\label{strongcollapse}
\end{figure}

The only question left to answer is how far the first region extends.
We see from Fig. 1 that for $L=50$ it covers roughly $70\%$ of the range
of $\phi$ above the threshold for single level selection, $\phi=1-\mu$,
which is surprisingly high.  The explanation of this depends on the
results of the next section, with the intermediate range asymptotics.
We anticipate the results by noting that the leading order correction
to $N(\phi)$ is, for $\ab=2$, $\delta N \sim {N}^2/L$.  
Thus, the first-region results start failing for $\delta N/N
 \approx 0.05$,
or $\bar{N}/L \approx 0.05$.
For large $L$, $\chi \sim 2\sqrt{N/L}$, we expect that for large
$L$ the first-region results are trustworthy out to $\chi \approx
2\sqrt{.05} \approx 0.45$, i.e.; $45\%$ of the region of interest.  
For smaller $L$, if fact things are better as we have seen.  It can be
shown that a more accurate criterion is
$(N-2.338N^{1/3})^2/(NL) \approx 0.05$. Applying this to $L=50$ gives
$N \approx 10$ which compares quite well with the numerical results in Fig. 1.
For $\nu\ne 1$, the first region is much narrower, though still a finite
fraction of the allowed range of $\chi$, independent of $L$.  
The first region approximation here
fails  when $\chi \approx 0.1\nu/\bar{\nu}$, which for $ab=4$, ($\nu=1/3$),
reads $\chi \approx .05$, very much smaller.  We can see this in Fig. 
\ref{firsta=4}, where again our approximation is over-conservative for
smaller $L$.

\begin{figure}
\centerline{
\includegraphics[width=4.0in]{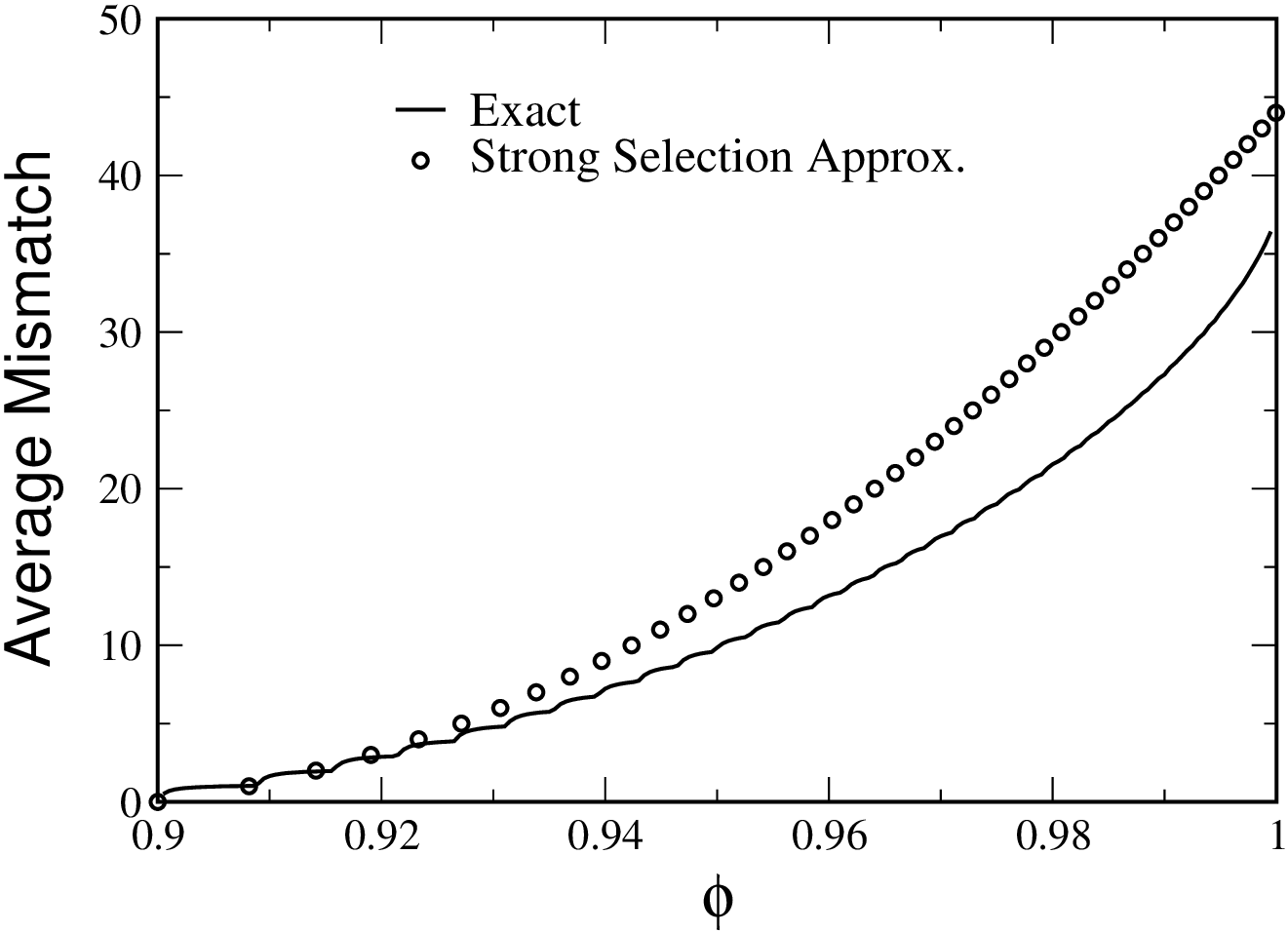}}
\caption{Comparison between exact results for $L=50$, $\mu=0.1$, $\ab=4$
and the
first-region approximation, Eq. \protect{\ref{1st_ans}}} 
\label{firsta=4}
\end{figure}

\section{Intermediate Selection, $N \gg 1$, $L/(1+\nu)-N \gg 1$}
\label{inter}

When $N$ gets too large, of order $L$, we can no longer
neglect the $\frac{\bar{\nu}(N-1)}{L}$ and $\frac{N-2}{L}$ terms
in the recursion relation for the determinants.
It is more convenient now to consider the $P_n$'s directly, using
the difference equation \ref{basiceq}.
Motivated by the result that, in the first region, the $P_k$'s grow 
rapidly with $k$, we assume that here too the same rule applies: 
The relevant $P$'s are those with $k \sim N$. Writing $N=\alpha L$, we 
thus assume that $k\sim\alpha L$. Plugging this into Eq. (\ref{basiceq}), 
we get the difference equation 
\begin{equation}
\chi P_k=\bar{\nu}\alpha P_k + \nu \alpha P_{k+1} + (1-\alpha)P_{k-1}
\label{Pn1}
\end{equation}
with the solution  
\begin{equation}
P_n=A\lambda_+^{n}+B\lambda_-^{n}
\label{solution1}
\end{equation}
where
\begin{equation}
\lambda_{\pm}=\frac{\chi - \bar{\nu}\alpha \pm
\sqrt{(\chi-\bar{\nu}\alpha)^{2}-4\nu\alpha(1-\alpha)}}
{2\nu\alpha}
\label{lambda}
\end{equation}

There are 3 options for $\lambda_+$ and $\lambda_-$: 1. They can both be real,
but this is incompatible with $P_{-1}=P_{N}=0$. 
2. They can both be complex, but then 
we get an oscillating solution,
which is also not acceptable, since the $P$'s must be non-negative. 3. The
 degenerate solution, where $\lambda_+=\lambda_-\equiv\lambda$, 
and the solution that satisfies $P_{N}=0$ is
\begin{equation}
P_n=A\lambda^{n}(N-n)
\label{solA}
\end{equation}
This solution indeed decays rapidly on a scale of order 1 as $n$ decreases
from $N$, justifying our initial assumption.  It also means that the
change in the solution from one $k$ to the next is also of order 1, and thus
the difference equation can not be approximated by a differential equation,
as PGHL attempted.

The condition for degeneracy of the $\lambda$'s is
\begin{equation}
\chi=\bar{\nu}\alpha + 2\sqrt{\nu\alpha(1-\alpha)}
\label{solutionchi}
\end{equation}
hence
\begin{equation}
\phi=\mu\chi+1-\mu=1-\mu + \mu\bar{\nu}\frac{N}{L} + 2\mu\sqrt{\nu\frac{N}{L}(1-\frac{N}{L})}
\label{phi}
\end{equation}

The last task is to calculate $\bar{k}$. Since the $P_n$'s decay exponentially
(with an $L$-independent width) as $n$ decreases,
to leading order in $1/L$, 
\begin{equation}
\bar{k}=N \ .
\end{equation}
Since here $N=\alpha L$, there is a finite probability, $q=\bar{k}/L$ 
of a mismatch per nucleotide in the intermediate selection region.
Combining results, we have
\begin{equation}
q=\frac{\bar{k}}{L}=\frac{1}{(1+\nu)^2}\left[2\nu+\chi\bar{\nu}-2\sqrt{\nu
(\nu + \bar{\nu}\chi-\chi^2)}\right]
\label{qbar}
\end{equation}
A comparison of the formula in Eq. (\ref{qbar}) and numerical results is 
given in 
Fig. (\ref{2ndgraph}) for $\ab=2$, and in Fig. (\ref{a4graph}) for $\ab=4$ 
($\nu=1/3$).  
We see that our formula overlaps that of the first region
for $\alpha \ll 1$, so that $\chi \approx 2\sqrt{\nu\alpha}=2\sqrt{\nu N/L}$,
which is indeed the first region result for large $N$. The leading correction
in the overlap can be calculated, leading to the results discussed above in
Sec. \ref{strong}. The second region solution has $q$ 
increasing from 0 at
$\chi=0$ ($\phi=1-\mu$) to the fully random value of $1/(1+\nu)=(\ab-1)/\ab$
at $\chi=1$ ($\phi=1$), so that this second region result covers the entire
range of $q$ and $\phi$.  Indeed, for fixed $\phi$, this result
is asymptotically correct for sufficiently large $L$.  At small $\chi$, our
result only breaks down when $N$ is not large, which only happens for
$\chi \sim O(L^{-1/2})$.  Similarly, it  breaks down $\alpha$ is too
close to $1/(1+\nu)$, since then $\lambda$ goes to unity and the
relevant $k$'s are no longer concentrated in a region around $k=N$.  
We shall see
in Sec. \ref{weak}, that this occurs for $1/(1+\nu) - \alpha$
of order $L^{-1/2}$, so that $\lambda$ is of order $\sqrt{L}$.  Thus, except
for small regions of $\phi$ near the very weak and very strong limits,
the size of which vanish in the large $L$ limit, our result Eq. (\ref{qbar})
is reliable.

\begin{figure}
\centerline{\includegraphics[width=4.0in]{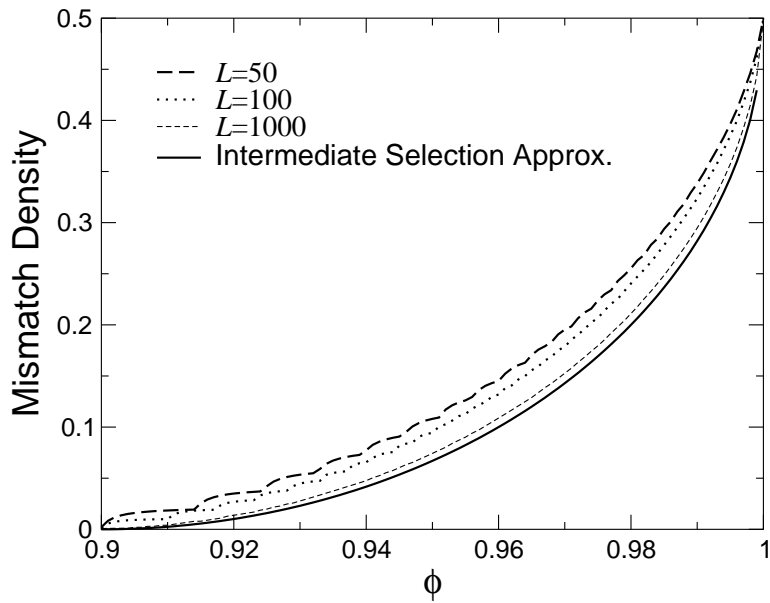}}
\caption{Comparison between the numerical results for the mismatch density,
$q$ vs. $\phi$  and Eq. (\protect{\ref{qbar}}) 
for $L=50$, 100, 1000 with $\mu=0.1$, $\ab=2$.}
\label{2ndgraph}
\end{figure} 

\begin{figure}
\centerline{
\includegraphics[width=4.0in]{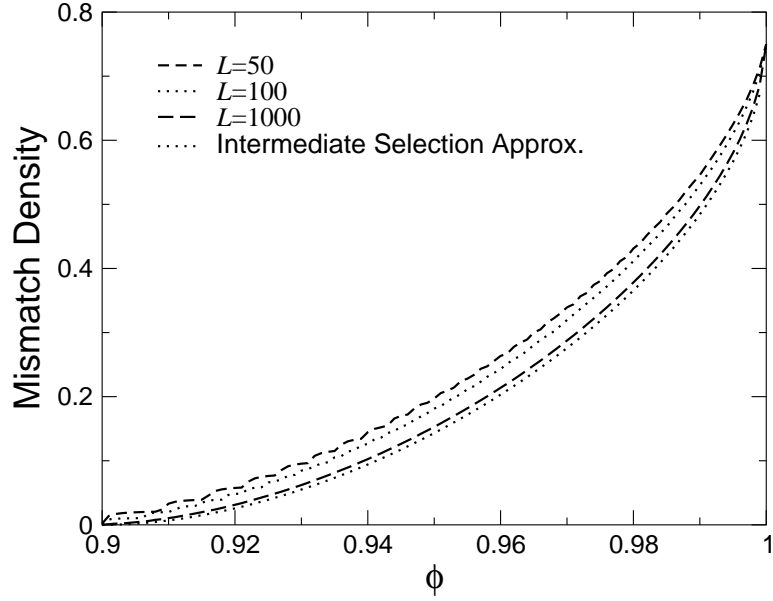}}
\caption{Comparison between the numerical results for the mismatch
density, $q$ vs. $\phi$ 
 and Eq. (\protect{\ref{qbar}}) for $L=50$, 100, 1000 with 
$\mu=0.1$, $\ab=4$. }
\label{a4graph}
\end{figure}

\section{Intermediate Selection, General $\mu$}
\label{inter1}

Up to now we have only considered the case of small mutation
rate, so that only single transitions had to be considered.  We now
take up the case of large $\mu$, dealing with multiple transitions
in a single round.  Again, we are working in the intermediate selection
region where only the states with $k \approx N=\alpha L$ are relevant.
We take the probability of a single site mutation $\mu_0$ to be small,
but $\mu=\mu_0 L$ to not necessarily be small, so that we may use the
Poisson distribution for the number of up and down mutations, considered
separately.
The mean number of up (increasing mismatch) mutations is $\mu(1-\alpha)$
and the mean number of down mutations is $\mu\nu\alpha$.
Consider the case, for example, 
where after the mutation
phase you wind up at the same level where you began.  This could result,
of course, from having no mutations of any kind, which has a probability 
$e^{-\mu(1-\alpha)}\cdot e^{\mu\nu\alpha} =e^{-\mu + \mu\alpha\bar{\nu}}$.
In addition, however, one can
wind up back where one started by having two mutations, one up and one
down.  The probability for this is 
$\left(e^{-\mu+\mu\alpha\bar{\nu}}\right) \cdot \mu(1-\alpha) \cdot 
\mu \nu\alpha$.
One can also return to the
original state after 4, 6, $\ldots$ mutations.  The total probability
of remaining in the original state is then
$$
e^{-\mu+\mu\alpha\bar{\nu}}\sum_{k=0}^{\infty}\frac{\mu^{2k}}{(k!)^2} 
\left(\alpha\nu(1-\alpha)\right)^k  =
e^{-\mu+\mu\alpha\bar{\nu}}I_0\left(2\mu\sqrt{\alpha\nu(1-\alpha)}\right)
$$
where $I_0$ is the modified Bessel function.
Henceforth, we shall write $2\mu\sqrt{\alpha\nu(1-\alpha)}\equiv t$.  Proceeding
likewise, the probability of mutating down from state $n+j$ ($j > 0$)
to state $n$ is 
$$
\left[\alpha\mu\nu\right]^j e^{-\mu+\mu\alpha\bar{\nu}}\sum_{k=0}^{\infty}\frac{\left(\alpha\nu(1-\alpha)\right)^k \mu^{2k}}{k!(k+j)!} =
e^{-\mu+\mu\alpha\bar{\nu}}\left(\frac{\alpha\nu}{1-\alpha}\right)^{j/2} I_j(t)
$$
and up from $n-j$ to $n$ is
$$
\left[(1-\alpha)\mu\right]^j e^{-\mu+\mu\alpha\bar{\nu}}\sum_{k=0}^{\infty}\frac{\left(\alpha\nu(1-\alpha)\right)^k \mu^{2k}}{k!(k+j)!} =
e^{-\mu+\mu\alpha\bar{\nu}}\left(\frac{\alpha\nu}{1-\alpha}\right)^{-j/2} I_j(t)
$$
Putting this all together yields the recursion relation ($I_{-j}=I_j$)
\begin{equation}
\phi P_n = e^{-\mu+\mu\alpha(1-\nu)} \sum_{j=-\infty}^{\infty} P_{n+j}\left(\frac{\alpha\nu}{1-\alpha}\right)^{j/2} I_j(t)
\end{equation}
The characteristic polynomial (here of infinite order) is
\begin{equation}
0 = -\phi + e^{-\mu+\mu\alpha(1-\nu)} \sum_{j=-\infty}^{\infty} \lambda^j\left(\frac{\alpha\nu}{1-\alpha}\right)^{j/2} I_j(t)
\end{equation}
This sum can be performed (see, e.g. \cite{Abrom_Stegun}, Eq. 9.6.33), yielding 
\begin{equation}
F(\lambda;\phi) = -\phi + e^{-\mu+\mu\alpha(1-\nu)} e^{\left(\frac{\lambda\sqrt{\alpha\nu}}{\sqrt{1-\alpha}}
+ \frac{\sqrt{1-\alpha}}{\lambda\sqrt{\alpha\nu}}\right)\frac{t}{2}}
 = 0 
\label{Flphi}
\end{equation}
The condition for degeneracy is $\frac{\partial F}{\partial \lambda}=0$,
yielding $\lambda=\sqrt{\frac{1-\alpha}{\alpha\nu}}$, exactly as in the
small $\mu$ case! Plugging this into Eq. (\ref{Flphi}) gives
\begin{eqnarray}
\phi &=& e^{-\mu+\mu\alpha\bar{\nu}+t} \\
&=&e^{\mu\left(2\sqrt{\alpha\nu(1-\alpha)}-1\right)+\alpha\mu\bar{\nu}}
\end{eqnarray}
In the limit of small $\mu$, $\phi\approx 1+\mu\left(2\sqrt{\alpha\nu(1-\alpha)}+\alpha(\bar{\nu}-1\right)$
which is of course what we got before. As with small $\mu$, the mismatch
density $q$ is just $\alpha$ to leading order, so that 
\begin{equation}
\phi = e^{\mu\left(2\sqrt{q\nu(1-q)}-1+q(1-\nu)\right)} .
\end{equation}
Turning this around gives
\begin{equation}
q=\frac{1+\nu+ \frac{\ln\phi}{\mu}\bar{\nu}-2\sqrt{
\nu\left(\frac{-\ln\phi}\mu\right)\left(1+\nu+\frac{\ln\phi}{\mu}\right)}}
{(1+\nu)^{2}}
\label{q0}
\end{equation}
A comparison between Eq. (\ref{q0}) and numerical results for $\nu=\frac{1}{3}$ and $\mu=1$ is shown in Fig.
(\ref{graph}).
\begin{figure}
\includegraphics[width=4.0in]{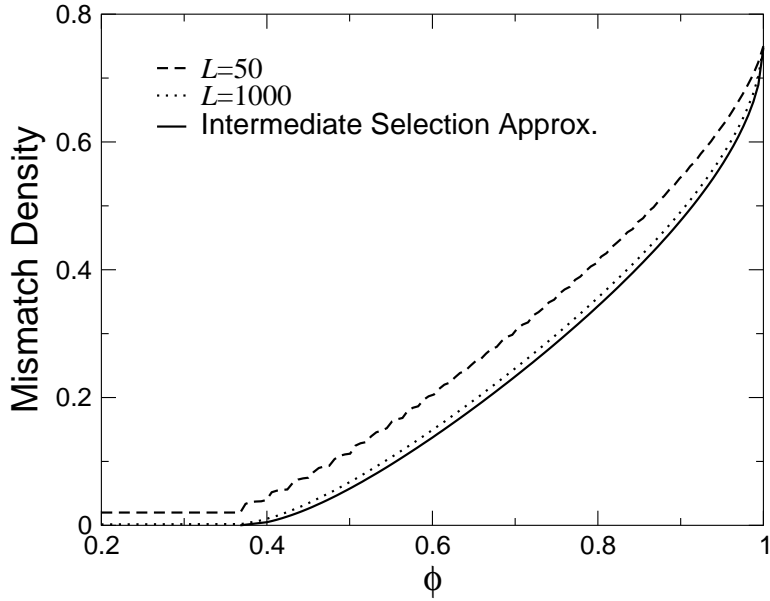}
\caption{Comparison between numerical results for the mismatch density
$q$ vs.  $\phi$ 
 and Eq. (\protect{\ref{q0}}) }
\label{graph}
\end{figure}
The formula for general $\mu$ can be seen to reproduce exactly that for small 
$\mu$ with the replacement $(1-\phi)$ by $-\ln(\phi)$, 
with the control parameter
in general being $-\ln(\phi)/\mu$, so that stronger mutation can always be
exactly compensated by stronger selection.

\section{Weak Selection}
\label{weak}

\subsection{Determining $\chi(N)$}

We return to Equation (\ref{basiceq}) describing the equilibrium state.
We saw from the second region
solution that what $N$ approaches $L/(1+\nu)$, the width of the distribution
diverges and one can no longer consider $k$ as a constant.
However, the fact that the distribution is very broad (of order
$\sqrt{L}$ as we will see) means that it changes
slowly with $k$, which allows us to expand $P_{k+1}$ and $P_{k-1}$ 
in a Taylor series, yielding:
\begin{eqnarray}
\left(1-\chi+\frac{1+\nu}{L}\right)P_x &+& 
\left((1+\nu)\frac{x}{L}-1 - \frac{\bar{\nu}}{L}\right)P'_x
\nonumber \\ {} &+& \frac{1}{2} \left(-\bar{\nu}\frac{x}{L} + 
1+\frac{1+\nu}{L}\right) P''_x = 0.
\label{px}
\end{eqnarray}
where $x$ is a continuous variable replacing the discrete variable $k$ in 
Eq. (\ref{basiceq}).  
We can eliminate $L$ by going over to the scaled variables
\begin{equation}
y\equiv \frac{x - \frac{L}{1+\nu}}{\sqrt{L}} \quad ; \quad \delta \chi \equiv L(1-\chi),
\end{equation}
yielding to leading order
\begin{equation}
\frac{\nu}{1+\nu} P''_y 
+\left((1+\nu)y\right)P'_y
+\left(\delta\chi + 1+\nu\right)P_y 
= 0.
\label{py}
\end{equation}
We again transform into a Schroedinger equation, via then
similarity transformation $P_y=f_y g_y$ with
\begin{equation} 
g_y=e^{-\frac{(1+\nu)^2 y^2}{4\nu}}
\label{g}
\end{equation}
yielding
\begin{equation}
-\frac{\nu}{1+\nu}f''_y + \left(\frac{(1+\nu)^3}{4\nu}y^2 - \frac{1+\nu}{2}
- \delta\chi\right)f_y = 0. 
\label{feq}
\end{equation}
Rescaling $z=(1+\nu)y/\sqrt{\nu}$, we
obtain (yet again) the Schroedinger equation for a harmonic oscillator,
\begin{equation}
-\frac{1}{2}f''_z + \frac{1}{8}z^2 f_z = Ef_z
\label{shrod}
\end{equation}
in which $m=\hbar=1$, $\omega=1/2$, with the energy term reading
\begin{equation}
E=\frac{1}{4} + \frac{\delta\chi}{2(1+\nu)}
\label{Eeqn}
\end{equation}
The boundary conditions, 
$P_{-1}=P_N=0$, are equivalent to infinite walls at $z=-\sqrt{L/\nu}$ and
$z=((1+\nu)N-L)/\sqrt{\nu L}$. 
Because the factor $g_z$ decays rapidly as $z$ decreases away from 0,
we can replace the left boundary condition by $P(z=-\infty)=0$.
 at the cost of an exponentially small error.
Thus, for a given $\chi$, or equivalently $E$, as we come in from $y=-\infty$ we want to find the first $z=z^*$ for which
$f_z^*=0$, which then determines $N$ through
\begin{equation}
N=\frac{L}{1+\nu}\left(1 + z^*\sqrt{\frac{\nu}{L}}\right) .
\label{Neqn}
\end{equation}
Thus, as we noted in the beginning, $N$ is a distance of order $\sqrt{L}$ from
$L/(1+\nu)$ in the third region.
There are of course a set of $E$'s for which we know $z^*$ analytically. If
$E=(n+\frac{1}{2})\omega$ for some whole number $n$, then the solution of
the Schroedinger equation is a Gaussian times the Hermite polynomial, 
$\textrm{He}_n$,
and so $z^*$ in this case is just the largest (negative) 
root of the $\textrm{He}_n$.  
It is remarkable that, as in the first region, the solution is determined by
the zeroes of the Hermite polynomial. However, whereas there the quantum
level $n$ increased with $\phi$, here it is just the opposite.
Now, the root $z^*$ decreases with $n$, so that as $\chi$ decreases from
1, $N$ decreases, as we expect. At $n=1$, the wave function has a single
root at the origin, so $z^*$ vanishes, and so $N=L/(1+\nu)$ at this point.
For the case of general $E$, the solution for
$f$ is a parabolic cylinder function, and the root $z^*$ has to be
determined numerically.  In particular, for $E<\frac{3}{2}\omega$,
$z^*$ is positive, so $N>L/(1+\nu)$. As the ground state wave function has no
zeros, $z^*$ must go off to $\infty$ as $E$ approaches the ground
state energy of $E=\frac{1}{2}\omega$.  In fact, the case 
$E=\frac{1}{2}\omega$ is
precisely the limiting case $\chi=1$ where there is no 
selection and so $N=L$.  The divergence we obtain is consistent with
with the fact that the exact $z^*$ is not order 1, but rather of order 
$\sqrt{L}$ at this
point, and is a result of our approximation of the left boundary condition.
Thus, the scaling variables in the third region are $E$ and $z^*$, which
are related to the physical variables $\chi$ and $N$ via Eqs. (\ref{Eeqn}) and
(\ref{Neqn}). 
The results for the numerical calculation of $z^*$ vs. $E$ is
presented in Fig. \ref{graph1}, together with the exact solution for
integer $n$ in terms of the zeros of the Hermite polynomial and
the asymptotic formula for large $E$ derived in the section on the
first region: $z^{*}=-(2\sqrt{n}-2.338n^{-\frac{1}{6}}+\frac{1}{2\sqrt{n}})\approx
-(2\sqrt{2E}-2.083E^{-1/6}$).

\begin{figure}
\includegraphics[width=4.0in]{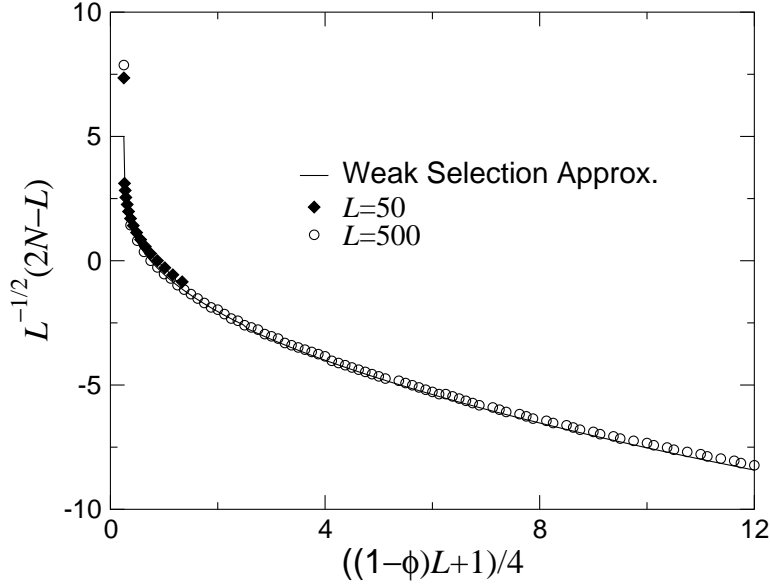}
\caption{Comparison between the numerical solution of {\protect{$z^{*}=(2N-L)/\sqrt{L}$}} 
vs. $E=((1-\phi)L+1)/4$
together with the scaled results of the exact numerical solution of
Eq. (\protect{\ref{exacteq}}) for $L=50$, 500, with $\mu=0.1$, $\ab=2$.}
\label{graph1}
\end{figure}

\subsection{Finding $\bar{x}(N)$}

Using the relation between the variables $x$ and $z$ above, 
we will calculate the average $\bar{x}$ via 
\begin{equation}
\bar{x}= \frac{L}{1+\nu} + \frac{\sqrt{\nu L}}{1+\nu}\bar{z}
\end{equation}
where
\begin{equation}
\bar{z}=\frac{\int_{-\infty}^{z^{*}} z
f_z e^{-\frac{z^2}{4}}dz}{\int_{-\infty}^{z^{*}}f_z e^{-\frac{z^2}{4}}dz} .
\label{integralz}
\end{equation}
where $z^*$ is as above the value of $z$ for which $f_z=0$.
Using the numerical solution for $f$ for various $E$'s allows us to construct
the scaling function $\bar{z}(E)$,
the scaled $\bar{k}$, (i.e. $\frac{(1+\nu)\bar{k}-L}{\sqrt{\nu L}}$) vs. the scaled selection parameter, $E=1/4 + (1-\chi)L/2(1+\nu)$ as shown in
Fig (\ref{scaledqbar}).

\begin{figure}
\includegraphics[width=4.0in]{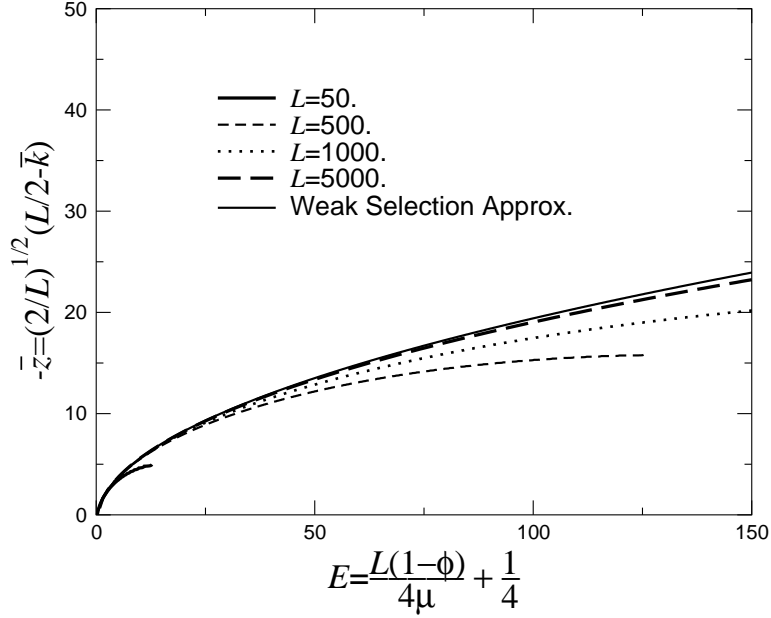}
\caption{Comparison between $\bar{z}$, the scaled average mismatch vs. $E$,
the scaled selection parameter from the numerical calculation of 
Eq. (\protect{\ref{integralz}}) and 
exact numerical results
for $L=50$, 500, 1000, and 5000, with $\mu=0.1$, $\ab=2$.}
\label{scaledqbar}
\end{figure}

We can generate an approximation to $\bar{z}$ for large $E$.  Here as
we saw, $z^*$ is large and negative and so
 the Gaussian factor in the integral
restricts the relevant range of $z$ to those near $z^*$, which
as we saw in the Region 1 calculation, is itself near the classical turning
point $z_0 \equiv -2\sqrt{2E}$.  Proceeding as in Sec. \ref{strong}, we
introduce  $z=z_0 + A\tilde{z}$, $A=(-z_0/2)^{-\frac{1}{3}}$, and requiring
$f(-\infty)=0$ yields
\begin{equation}
f(\tilde{z}) = \textrm{Ai}(-\tilde{z}) .
\end{equation}
We do not worry about normalization since we explicitly
divide by the normalization integral.
This leads to 
\begin{equation}
\bar{z}=\frac{\int_{-\infty}^{\tilde{z}^*} (z_0+A\tilde{z})
\textrm{Ai}(-\tilde{z}) e^{-z_0A\tilde{z}}d\tilde{z}}{\int_{-\infty}^{\tilde{z}^*} 
\textrm{Ai}(-\tilde{z}) e^{-z_0A\tilde{z}}d\tilde{z}} .
\label{integralairy}
\end{equation}
where $\tilde{z}^*=2.338$ is the first zero of $f$.
Changing the integration variable $\tilde{z}$ to $s=\tilde{z}^*-\tilde{z}$
yields: 
\begin{equation}
\bar{z}=-z_0-A\tilde{z}^* - \frac{\int^{\infty}_{0}e^{z_0A s/2}s \textrm{Ai}(s-\tilde{z}^*)
ds}{\int^{\infty}_{0}e^{z_0A s/2} \textrm{Ai}(s-\tilde{z}^*) ds} .
\label{ints}
\end{equation}

Because of the fast decay  of the exponential factor, ($|z_0| \gg 1$), 
the main contribution for the integral in (\ref{ints}) is around $s=0$
This allows us to expand $\textrm{Ai}(s-\tilde{z}^*)$ in (\ref{ints}) in
a  Taylor series: $\textrm{Ai}( s-\tilde{z}^*)\approx s \textrm{Ai}'(-\tilde{z}^*)+\ldots$.
This gives:
\begin{eqnarray}
\bar{z}&=&z_0 + A\tilde{z}^* - \frac{\int_0^\infty  s^2 e^{z_0As/2}ds}
{\int_0^\infty  s e^{z_0As/2}ds} = 
z_0 + A\tilde{z}^* - \frac{4}{Az_0} \nonumber \\
&=& -2\sqrt{2E} + 2.3338\left(2E\right)^{-1/6} - \left(E/4\right)^{-1/3} .
\label{solution}
\end{eqnarray}
On the scale of Fig. (\ref{scaledqbar}), the difference between this formula
and the exact numerical calculation is not visible.

\section{Next-Order Correction to Intermediate Selection}
\label{inter2}

In order to better understand the origin of the degeneracy condition
we invoked to solve the second region problem, it is worthwhile to
explore the next-order correction.  In addition, we will be able to
understand the relatively poor accuracy of our result for
$\bar{q}(\phi)$.  We will for simplicity restrict ourselves in this
section to the case $\ab=2$.  In the second region, we could not replace the
difference equation by a differential equation, since the change in $P$
from step to step was of order 1.  We will fix this by rescaling $P$
by the geometrical series behavior we found, by putting $Q_n=\frac{P_n}{\lambda^n}$, where $\lambda$ is as determined above, $\lambda=\sqrt{(1-\alpha)/\alpha}$,
so that the differential equation:
\begin{equation}
\chi P_n=(1-\frac{n-1}{L}) P_{n-1}+\frac{n+1}{L} P_{n+1}.
\label{eq1}
\end{equation}
becomes
\begin{equation}
{\chi}Q_n \lambda^n=Q_{n-1}\lambda^{n-1}(1-\frac{n-1}{L})+Q_{n+1}\lambda^{n+1} \frac{n+1}{L}.
\label{Qn}
\end{equation}
We now assume the $Q$'s vary smoothly and expand $Q_{n-1}$ and $Q_{n+1}$ 
in a Taylor series, (writing $n=\alpha L + y$), and dropping all
terms of order $L^{-1}$, though not $y/L$) yielding
\begin{equation}
\left[\chi_0 + \frac{\beta y}{L}\right]\frac{Q''}{2}
+ \left(\frac{\gamma y}{L}\right)Q' + \left[\chi_0 - \chi + 
\frac{\beta y}{L}\right] = 0
\end{equation}
where $\chi_0=\lambda\alpha + (1-\alpha)/\lambda=2\sqrt{\alpha(1-\alpha)}$, $\gamma=\lambda+1/\lambda=2/\chi_0$ and $\beta=\lambda-1/\lambda=(1-2\alpha)\gamma$. 
As usual, we transform into a Schroedinger equation by a similarity
transformation, $Q=fg$, with:
\begin{equation}
g=\left(\chi_0 + \frac{\beta y}{L}\right)^{\frac{\chi_0\gamma L}{\beta^2}}
e^{-\frac{\gamma y}{\beta}}
\label{gx}
\end{equation}
giving
\begin{equation}
-\frac{1}{2} f''+ f\left[\frac{\chi-\chi_0 - \frac{\beta y}{L}}
{\chi_0 + \beta y/L}
+ \frac{\gamma^2 y^2}{2L^2 (\chi_0 + \beta y/L)^2}\right]=0
\label{shro}
\end{equation}
where we have again dropped a term of order $L^{-1}$.
We see that the coefficient of $f$ in Eq. (\ref{shro}) vanishes
at $y=0$ if $\chi$ is taken equal to $\chi_0$, the zeroth order solution.
Thus the zeroth order solution corresponds to choosing the ``energy''
$-\chi$ so that the classical turning point is at
$y=0$, or $n=N$.  However, the true boundary condition is that $f$ vanish
at $n=N$, which happens some small distance (relative to the
length scale for changes of the potential $L$) behind the
turning point.  Thus, we must choose $\chi$ slightly less
than $\chi_0$ in order to move the turning point to negative $y$.
Assuming $y/L$ and $\delta{\chi}\equiv \chi-\chi_0$ are small, we get an Airy equation:
\begin{equation}
-\frac{1}{2} f''+ f\left[\frac{\delta\chi}{\chi_0} - \frac{\beta y}{
\chi_0 L}\right]=0
\label{shro1}
\end{equation}
The solution of this equation that decays for negative $y$ is 
\begin{equation}
f=\textrm{Ai}\left(-\left(\frac{2\beta}{\chi_0 L}\right)^{1/3}
\left(y-\frac{L \delta\chi}{\beta}\right)\right)
\label{airy}
\end{equation}
The requirement that the first zero of $f$ lie at $y=0$ implies
\begin{equation}
\delta{\chi}=\frac{-2.338\beta^{2/3}(\chi_0/2)^{1/3}}{L^{2/3}} =
\frac{-2.338(1-2\alpha)^{2/3}}{(a(1-\alpha))^{1/6}L^{2/3}} 
\label{deltachi}
\end{equation}

A comparison of the formula in (\ref{deltachi}) and the exact numerical
results is given in Fig. (\ref{dchifig}).
\begin{figure}
\centerline{
\includegraphics[width=4.0in]{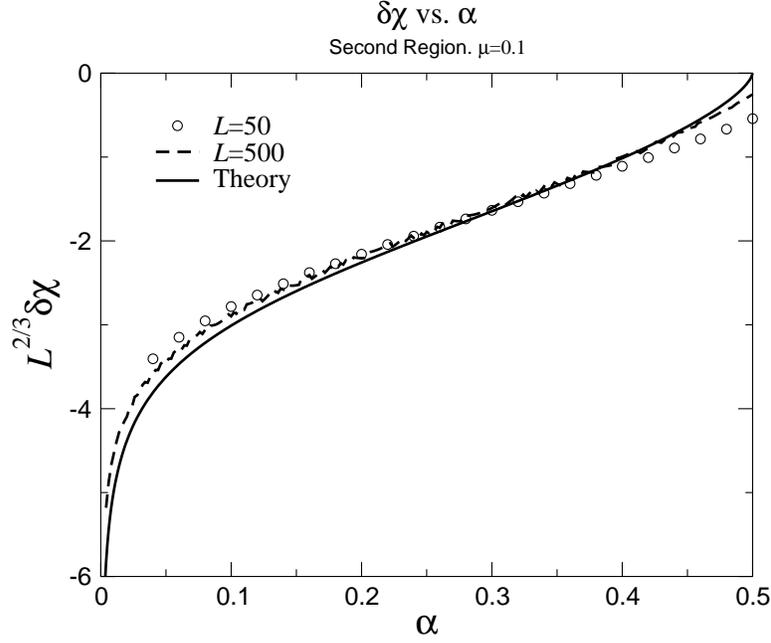}}
\caption{Comparison between the exact numerical results for 
$\delta\chi$ vs $\alpha$ 
 and the formula in Eq. (\protect{\ref{deltachi}}) }
\label{dchifig}
\end{figure}

Remembering that $\chi+\delta\chi=\frac{\phi-1+\mu}{\mu}$ yields:
\begin{equation}
\phi=1-\mu + \frac{\mu}{L}\sqrt{N(L-N)}\left[2-2.338\left(\frac{1}{N}-\frac{1}{L-N}\right)^{\frac{2}{3}}\right]
\label{phiN}
\end{equation}
where $N=\alpha L$. 
Now, let's remember we got for the second region at zeroth order
$P_n \propto (N-n)\lambda^{n}$. We can calculate $\bar{n}$ as follows:
\begin{equation}
\bar{n}=\frac{\sum_{n=0}^{\alpha L}{nP_n}}{\sum_{n=0}^{\alpha L}{P_n}}.
\label{nbar}
\end{equation}
with $N=\alpha L$, and we get, up to exponentially small terms
\begin{equation}
\bar{n}=N-\frac{L+2\sqrt{N(L-N)}}{L-2N}.
\label{solnbar}
\end{equation}
so we get a correction for $\bar{n}-N$. Combining this correction with the 
correction for
$\phi$ in (\ref{phiN}), We get a more accurate graph of $\bar{n}$ as 
a function of $\phi$.

\begin{figure}
\centerline{
\includegraphics[width=4.0in]{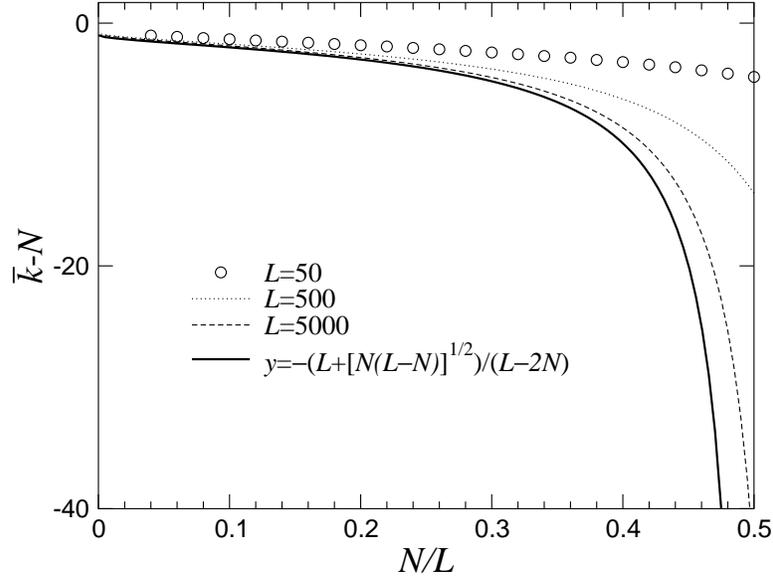}}
\caption{Comparison between the exact numerical results for 
$\bar{x}-N$ vs $\alpha$ 
and the formula in Eq. (\protect{\ref{solnbar}}) for $L=50$, $500$, and $5000$,
with $\mu=0.1$.}
\label{xbaralf}
\end{figure} 
\begin{figure}
\centerline{
\includegraphics[width=4.0in]{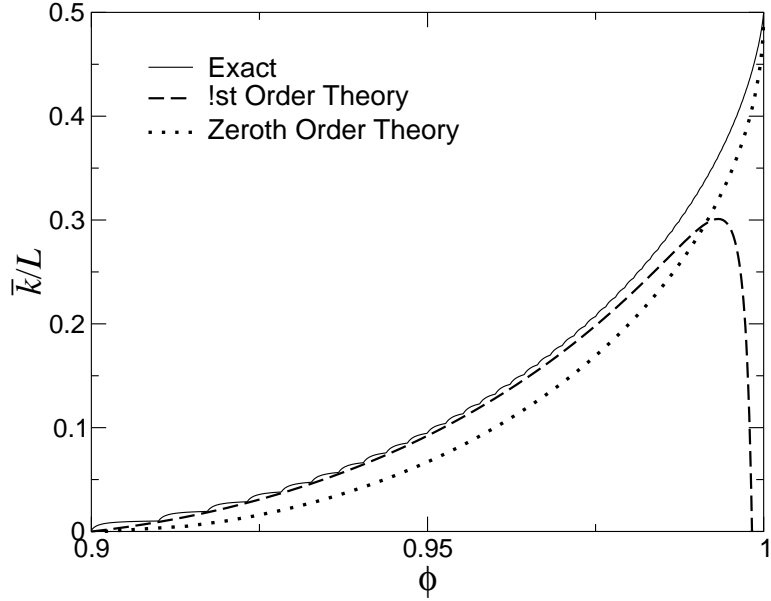}}
\caption{Comparison between the exact numerical results for $\bar{x}$ vs $\phi$ 
 and the formulas in Eqs. (\protect{\ref{phiN}}) and 
(\protect{\ref{solnbar}}) for $L=100$, $\mu=0.1$ $\ab=2$. }
\label{xbarphi}
\end{figure} 

\section{Comparison to PGHL}
\label{compare}
Before concluding, we wish to make some comments on the connection 
of this analysis to that
attempted by PGHL.  As we have noted, their approach was doomed by the
assumption of a continuum limit for the distribution function.  This is
true, as we have seen, only in the weak selection region, which is valid
only in a very limit range of $\phi$ near 1, of order $\mu/L$.  
Even there, their answer as stated is incorrect, due to their neglect of the
dependence of the effective diffusion constant on the local mismatch.  If
one uses for the diffusion constant that obtained at the {\em average}
mismatch, one obtains an implicit equation for the average mismatch at
equilibrium, which agrees with our results for the overlap region between
the second and third regions.  Neglecting this dependence forced PGHL
to introduce a parameter $\tau$ which they fitted numerically to
in essence restore the correct diffusion constant. Achieving agreement
with the asymptotic behavior for weak selection would have dictated a
value of $\tau=2$ in the infinite $L$ limit.  Fitting $\tau$ at the
finite value $L=170$ gave them instead a value of $\tau=2.77$.
Actually, their formula neglecting the mismatch dependence of the diffusion
constant and using $\tau$ actually results in better agreement with the
exact results, this agreement is the fortuitous outcome of a partial cancellation of the errors induced both by the mismatch dependence of the diffusion 
constant and the continuum approximation.  At strong selection, their
approach in any case is unreliable.
This situation is summarized in Fig. (\ref{herbfig}),
where we show the results of our theory together with the 
stated predictions of PGHL, along with the (partially) 
corrected form of their theory taking account of the mismatch
dependence of the the diffusion constant.
case they considered.  We see that the partially corrected form of PGHL
agrees near $\phi=1$ and subsequently systematically deviates from the
simulations.  The $\tau=2.77$ original formula numerically does better
overall, but is nowhere correct.
Nevertheless, the basic mechanism responsible for
selecting the equilibrium state, namely the degeneracy condition discussed
in Sec. \ref{inter}, was correctly identified by PGHL.  This degeneracy
condition is intimately connected with the marginally stable nature of the
dynamic problem\cite{Brunet}.

\begin{figure}
\centerline{\includegraphics[width=4.0in]{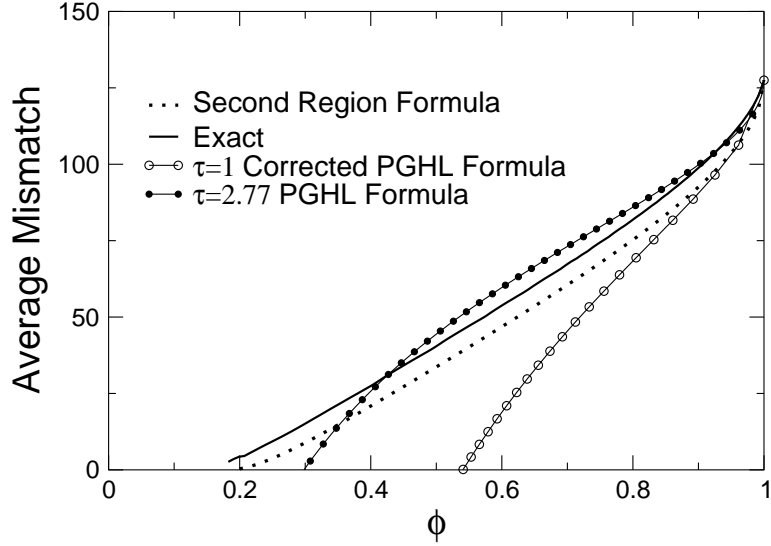}}
\caption{Comparison of the numerical results for the steady-state
for the parameters investigated by PGHL, $L=170$, $\mu=1.7$, $\ab=4$ and
our second-region theory, Eq. ({\protect{\ref{q0}}}) with the results
of the PGHL theory, with $\tau=2.77$ and the corrected version of
the PGHL theory, with the mismatch dependent diffusion constant and
$\tau=1$.}
\label{herbfig}
\end{figure}

\section{Conclusions}
\label{end}

We have seen that the steady-state equation is quite rich in its behavior,
with different behavior in each of the three separate regimes of the selection
parameter $\phi$.  As a summary, we present in Fig. \ref{summ_200} a graph
with all three approximations displayed, as well as the exact numerical 
solution for $N=200$, $\mu=0.1$, $\ab=2$.  We see that, as already
noted,  the strong 
selection approximation covers roughly the first half of the nontrivial 
range of $\phi$, while
the second-order intermediate selection result works essentially from
the smallest $\phi=1-\mu$ up to $\phi$ of 0.99.  The
weak selection theory works in this case already from $\phi=0.97$.

\begin{figure}
\includegraphics[width=4.0in]{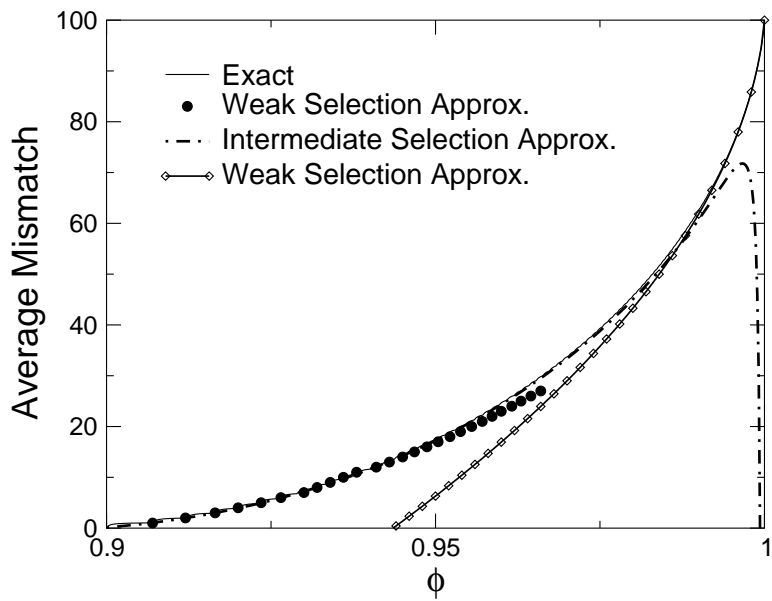}
\caption{Average mismatch vs. selection parameter $\phi$ for $L=200$, 
$\mu=0.1$, $\ab=2$, together with the three different approximations
appropriate for weak, intermediate and strong selection.}
\label{summ_200}
\end{figure}

A few qualitative features stand out from the solution and are worth
pointing out.  First, the average mismatch is generically a finite fraction
of the sequence length, except for very strong selection, extremely near
the threshold for selecting only the best state. Second, the distribution
of mismatches is very skewed toward high mismatch.  The most probable state
is very near the bottom of the barrel of the states that survive at all.  The
high affinity states are exponentially rare in the equilibrium distribution.
Signs are this behavior are perhaps discernible in the experimental data
of Ref. \cite{dubertret}.  Last is the exact tradeoff between selection
strength and mutation rate, due to the existence of a single lumped control 
parameter.  This opens the possibility for optimizing the breeding process
by increasing the mutation rate to speed the approach to equilibrium, at
no cost in the quality of the final distribution of affinities.  There is
of course a limit to this imposed by the finite population size, which we
have not considered at all in this work.  If the selection criteria is too
rigid, the finite population size will eventually begin to play a role.

\begin{acknowledgments}
The authors acknowledge the support of the Israel Science Foundation.
The authors thank Herbert Levine for discussions.
\end{acknowledgments}

\end{document}